\documentstyle[twoside]{article}

\def\d{\partial}
\def\dh{\mathop{\vphantom{\odot}\hbox{$\partial$}}}
\def\dl{\dh^\leftrightarrow}
\def\sqr#1#2{{\vcenter{\vbox{\hrule height.#2pt\hbox{\vrule width.#2pt
height#1pt \kern#1pt \vrule width.#2pt}\hrule height.#2pt}}}}
\def\w{\mathchoice\sqr45\sqr45\sqr{2.1}3\sqr{1.5}3\,}

\def\=d{\,{\buildrel\rm def\over =}\,}

\def\i3p{\p32\int d^3p}

\def\As{A\hbox to 1pt{\hss /}}
\def\np4{\int d^4p_1\cdots d^4p_{n-1}\, }

\def\nx4{\int d^4x_1\ldots d^4x_n\, }

\def\kon#1#2{\vbox{\halign{##&&##\cr
\lower4pt\hbox{$\scriptscriptstyle\vert$}\hrulefill &
\hrulefill\lower4pt\hbox{$\scriptscriptstyle\vert$}\cr $#1$&
$#2$\cr}}}
\def\lra{\longleftrightarrow}
\def\konv#1#2#3{\hbox{\vrule height12pt depth-1pt}
\vbox{\hrule height12pt width#1cm depth-11.6pt}
\hbox{\vrule height6.5pt depth-0.5pt}
\vbox{\hrule height11pt width#2cm depth-10.6pt\kern5pt
      \hrule height6.5pt width#2cm depth-6.1pt}
\hbox{\vrule height12pt depth-1pt}
\vbox{\hrule height6.5pt width#3cm depth-6.1pt}
\hbox{\vrule height6.5pt depth-0.5pt}}
\def\konu#1#2#3{\hbox{\vrule height12pt depth-1pt}
\vbox{\hrule height1pt width#1cm depth-0.6pt}
\hbox{\vrule height12pt depth-6.5pt}
\vbox{\hrule height6pt width#2cm depth-5.6pt\kern5pt
      \hrule height1pt width#2cm depth-0.6pt}
\hbox{\vrule height12pt depth-6.5pt}
\vbox{\hrule height1pt width#3cm depth-0.6pt}
\hbox{\vrule height12pt depth-1pt}}

\def\konw#1#2#3{\hbox{\vrule height12pt depth-1pt}
\vbox{\hrule height12pt width#1cm depth-11.6pt}
\hbox{\vrule height6.5pt depth-0.5pt}
\vbox{\hrule height12pt width#2cm depth-11.6pt \kern5pt
      \hrule height6.5pt width#2cm depth-6.1pt}
\hbox{\vrule height6.5pt depth-0.5pt}
\vbox{\hrule height12pt width#3cm depth-11.6pt}
\hbox{\vrule height12pt depth-1pt}}

\def\eh{{\scriptstyle{1\over 2}}}

\def\i{{\rm int}}

\def\m3{{\mu_1\mu_2\mu_3}}

\def\p{{(+)}}

\begin{document}

\thispagestyle{empty}
{\bf Z\"urich-University-Preprint ZU-TH-21/95}\\

\vskip0,5cm
\vbox to 2,5cm{ }
\centerline{\Large\bf A Note on Slavnov-Taylor Identities }
\vskip 0.3cm
\centerline{\Large\bf in the Causal Epstein-Glaser Approach }
\vskip2.0cm
\centerline{\large\bf Tobias Hurth$^*$}
\vskip 0.5cm
\centerline{\large\it
Institut f\"ur Theoretische Physik der Universit\"at Z\"urich}
\centerline{\large\it Winterthurerstr. 190, CH-8057 Z\"urich, Switzerland}
\vskip 2,5cm
{\bf Abstract.} -  An alternative approach to perturbative Yang-Mills theories
in 3+1 dimensional space-time based on the causal Epstein-Glaser method in QFT
was recently proposed.\\
In this short note we show that the set of identities
between C-number distributions expressing nonabelian gauge invariance in
the causal approach imply identities which are analogous to the well-known
Slavnov-Taylor identities. We explicitly derive the Z-factor relations
at one-loop level.

\vskip 0.5cm
{\bf PACS.} 11.10 - Field theory, 12.35C-General properties of quantum
chromodynamics.
\vskip 0.5cm
$^*)$ Emailaddress: hurth@physik.unizh.ch\\
$^*)$ Supported by the Swiss National Science Foundation\\
\vskip 0.3cm

\vfill\eject

\vskip1.5cm

Nonabelian Gauge Invariance was recently analysed in the causal Epstein-Glaser
approach to perturbative QFT  [1].
In this approach the S-matrix is directly constructed in the Fock space of the
free asymptotic fields in the form of a formal power series
$$S(g)=1+\sum_{n=1}^{\infty}{1\over n!}\int d^4x_1...
d^4x_n\,T_n(x_1,...,x_n)g(x_1)...g(x_n),\eqno(1)$$
where $g(x)$ is a tempered test function which switches the interaction. The
central objects are the n-point distributions $T_n$ which may be viewed as
mathematically well-defined time-ordered products.  The defining equations of
the theory in the causal formalism are the fundamental (anti-)commutation
relations of the free field operators, their dynamical equations and the
specific coupling of the theory $T_{n=1}$. The n-point distributions $T_n$ in
(1) are then constructed inductively from the given first order $T_{n=1}$
according to the Epstein-Glaser construction [2] which allows a direct
construction of the renormalized (finite) perturbation series without any
intermediate modifications. The physical infrared problem is naturally
separated by
the adiabatic switching of the S-matrix $S(g)$ with a tempered test function
$g$. \\
 Considering Yang-Mills theory  in four space-time dimensions, the
corresponding specific coupling in the Feynman gauge is
$$T_1=igf_{abc}({1\over 2}:A_{\mu a}A_{\nu b}F^{\nu \mu}_c:
-:A_{\mu a}u_b\d^\mu\tilde u_c :+$$
$$+ \alpha\quad \d_\mu(:u_a \tilde u_b A^{\mu}_c:) - \beta  :\d_\mu A^\mu_a
\tilde u_b u_c:)  \eqno(2)$$
where $\alpha$ and $\beta$ are free constants.
All field operators herein are well-defined free fields and these are the only
quantities appearing in the whole theory. The double dots denote their normal
ordering. The specific coupling $T_{n=1}(x)$ of the theory does not contain
quadrilinear terms proportional to $g^2$, the four-gluon-vertex nor the
four-ghost-vertex. Both terms are automatically generated in second order by
our gauge invariance condition [3].\\
$A_{\mu a}(x)$ are the (free) gauge potentials satisfying the commutation
relations (Feynman gauge)
$$[A_a^{(-)\mu}(x),A_b^{(+)\nu}(y)]=i\delta _{ab}
g^{\mu\nu}D_0^{(+)}(x-y),\eqno(3)$$
where $A^{(\pm)}$ are the emission and absorption parts of $A$ and
$D^{(\pm)}_0$ the (mass zero) Pauli-Jordan distributions. $u_a(x)$ and $\tilde
u_a(x)$ are the free massless fermionic ghost fields fulfilling the
anti-commutation relations
$$\{u_a^{(\pm)}(x),\tilde u_b^{(\mp)}(y)\}= -i\delta
_{ab}D_0^{(\mp)}(x-y).\eqno(4)$$
$f_{abc}$ denotes the usual antisymmetric structure
constants of the gauge group $SU(N)$. The time-dependence of $A,u$ and $\tilde
u$ in Feynman gauge is given by the wave equation
$$\w A_a^\mu(x)=0,\quad \w u_a(x)=0,\quad \w \tilde
u_a(x)=0, \eqno(5)$$
We define
$$F_a^{\mu\nu}\=d \d^\mu A_a^\nu -\d^\nu A_a^\mu. \eqno(6)$$
\\
\\
Now one considers the linear (abelian!) BRS transformations of the free
asymptotic field operators. The generator of the abelian operator
transformations is the charge
$$Q\=d \int d^3x\,(\d_{\nu}A_a^{\nu}{\dl}_0u_a), \quad Q^2=0,
\eqno(7)$$
with the (anti-)commutation relations
$$[Q,A_\mu^a]_-=i\d_\mu u_a,\quad \{ Q,\tilde{u}_a \}_+ = -i\d_\nu
A_a^\nu,\quad \{ Q,u_a \}_+ =0, \quad [Q,F_{\mu \nu}^a]_-=0. \eqno(8)$$

In addition to the charge $Q$, one defines the ghost charge
$$Q_c:=i \int d^3x :(\tilde u \stackrel{\leftrightarrow}{\d}_0 u): \eqno(9)$$
In the algebra, generated by the fundamental field operators, one introduces a
gradation by the ghost number $G (\hat A)$ which is given on the homogenous
elements by
$$[Q_c, \hat A]= - G(\hat A) \cdot \hat A. \eqno(10)$$
One can define an anti-derivation $d_Q$ in the graded algebra by
$$d_Q \hat A:= Q \hat A-(e^{i\pi Q_c} \hat A e^{-i\pi Q_c}) Q\eqno(11)$$
The anti-derivation $d_Q$ is obviously homogenous of degree (-1) and satisfies
$d_Q^2 = 0$.\\
\\
Nonabelian gauge invariance in the causal approach means that the commutator of
the specific coupling (2) with the charge $Q$ is a divergence (in the sense of
vector analysis):
$$d_Q T_{n=1}= i \d_\nu [igf_{abc} (:A_\nu^a u_b F_c^{\nu \mu} :- \eh : u_a u_b
\d^\nu \tilde{u}_c: - \alpha i d_Q(:u_a \tilde u_b A^\nu_c:) ) ] \=d i\d_\nu
T_{1/1}^\nu \eqno(12)$$
The second term in (2) (the gluon-ghost-coupling) is essential that
 $d_Q T_{n=1}$ can be written as a divergence. Note the  different compensation
of terms in the invariance equation (2) compared with the invariance of the
Yang-Mills Lagrangean under the full BRS-transformations of the interaction
fields in the conventional formalism. \\
$T_{n=1}$ in (1) represents the most general gauge invariant (in the sense of
(12)) and Lorentz invariant operator, which is also invariant in regard to the
global $SU(n)$ group and in regard to the discrete symmetry transformations C,
P, T , and which has maximal
mass dimension four and ghost charge zero. Note that terms with
four operators are ruled out by the gauge invariance condition (12). As we have
already mentioned above, the four-gluon- and four-ghost-couplings are
automatically generated by the gauge invariance condition in the second order.
Moreover,  we left out all possible gauge invariant terms with two operators
because the information of quadratic terms are already contained  in the
fundamental (anti-)commutation relations and the dynamical equations of the
operators. \\
$T_{n=1}$ in (2) is also anti-gauge invariant in the
sense that
$$ [ \bar Q , T_{n=1}] \quad ( \mbox{where} \quad \bar Q := \int
d^3x\,(\d_{\nu}A_a^{\nu}{\dl}_0 \tilde u_a) \quad \mbox{with} \quad \bar Q^2=0
) \quad $$
is also a divergence (in the sense of vector analysis).\\

\vskip1cm
The condition of nonabelian operator gauge invariance in the causal approach is
expressed in every order of perturbation theory separately by a simple
commutator relation of the n-point distributions $T_n$ with the charge $Q$, the
generator of the free operator gauge transformations:
$$[Q,T_n(x_1,...,x_n)]= d_Q  T_n(x_1,.....,x_n)= i\sum_{l=1}^n\d_\mu^{x_l}T^
\mu_{n/l}(x_1,...,x_n),\eqno(13)$$
where $T_{n/l}^\nu (x_1, \ldots, x_n)$ are n-point distributions of an extended
theory which contains, in addition to the usual Yang-Mills couplings
$T_{n=1}(x)$ (2), the so-called $Q$-vertex $T_{1/1}^\nu (x)$ which already
occurs in (11) as a divergence-representation of [$Q, T_1$].
The first order S-matrix of the extended theory is equal to
$$S_1(g_0,g_1)\=d \int d^4x \,[T_1(x)g_0(x)+
T_{1/1}^\nu (x)g_{1\nu}(x)].\eqno(14)$$
and $g_1=(g_{1\nu})_{\nu=0,1,2,3}\in ({\cal S}({\bf R}^{\it 4}))^4$ must be an
anti-commuting C-number field. The higher orders are determined by the usual
inductive Epstein Glaser construction up to local normalization terms.
The $T^\mu_{n/l}$ are the n-point
distributions of the extended theory with one $Q$-vertex at $x_l$, all other
$n-1$ vertices are ordinary Yang-Mills vertices (2) (for details see [1]).\\
The representation of $[Q, T_{n=1}]$ as a divergence is in general not unique.
The most general $Q$-vertex $\tilde T_{1/1}^\nu$ with the same mass dimension
and ghost number as $T_{1/1}^\nu$ in (12) is the following:
$$[Q,T_1]=i\d_\nu [T_{1/1}^\nu + \gamma B_{1/1}^\nu] \=d i \d_\nu
\tilde{T}_{1/1}^\nu$$
$$\mbox{with} \qquad B_{1/1}^\nu=igf_{abc} \d_\mu (: u_a A_b^\mu A_c^\nu:),
\quad \d_\nu B_{1/1}^\nu =0, \quad \gamma \in \mbox{C} \quad \mbox{free.}
\eqno(15)$$
The choice of $\gamma$ has just practical reasons and has no physical
consequences.
\\
\\
\\\\
We claimed in [1]  that the simple operator condition (13) involving only
well-defined asymptotic field operators expresses the full content of the
nonabelian gauge structure of the quantized theory. We  proved this condition
by induction on the order n of perturbation theory following the causal
construction of $T_n$ and $T_{n/l}^\nu$. We also proved that this condition
implies the unitarity of the S-matrix in the physical subspace , i.e. the
decoupling of the unphysical degrees of freedom in the theory.
Thus, the concept of abelian gauge transformations of the free field operators
is sufficient in order to derive the most important consequence of nonabelian
gauge invariance in perturbative quantum field theory.\\
However, one may doubt if this simple operator equation represents the
whole content of nonabelian gauge invariance in perturbation theory, for
example the consequences for (amputated) Greensfunctions, namely the
Slavnov-Taylor identities.
The purpose of this note is to show that the equation (13) also contains
this latter information.\\
In [1] we expressed the operator gauge invariance condition by a set of
identities between C-number distributions. These C-number identities for gauge
invariance (so-called cg-identities) are sufficient for the operator condition
(13).
We have rewritten these identities in the appendix. Note that they correspond
to the specific choice $\alpha=0$ and $\beta=0$  in (2), which corresponds
to the Faddeev Popov specific coupling. Moreover, we have chosen $\gamma=0$,
the most suitable choice for the Q-vertex.\\
It is an advantage of the causal approach that the physical infrared problem
is naturally separated by adiabatic switching of the S-matrix by a tempered
testfunction $g$ and also absent before the limit $g \rightarrow 1$ is taken.
So all examinations regarding gauge invariance and unitarity are mathematically
well-defined. But one has to pay a prize: In order to express the operator
gauge invariance condition (13) in a set of identities between C-number
distributions,
we have to work out the explicit form of the divergence in (12) and
(13)([1],[4]). Moreover, one has to distinguish the operator and its derivative
, which implies the relative largeness of the set of cg-identities.\\

However, we can derive 5 summed identities from this large set of identities
which are totally analogous to the Slavnov-Taylor identities.\\
In fact, we can eliminate all distributions with one $Q$-vertex besides the
divergences in regard to the inner variables. One arrives at relations which
almost only involve distributions of the orginal theory:
 In order to get the summed 2-leg identity, one has to insert (A.1) into (A.2).
Besides the 3-leg identity (A.3) one attends another summed (3-leg) identity
by inserting (A.6) into (A.5), then (A.5) into (A.4). The 4-leg identities are
treated analogously: Inserting (A.12) into (A.8), we get the first summed 4-leg
identity and inserting (A.11) into (A.10), then (A.10) into (A.9) and finally
(A.9) into (A.7), we arrive at the second summed 4-leg identity:\\

$\bullet$ In the first step we define the following summed distributions. These
definitions are natural, because the defined distributions represent in each
case the sum of all
distributions which would contribute to the same operator in the adiabatic
limit (Partial integrating is formally possible in the adiabatic limit.). The
crucial point is that also the four-gluon terms proportional to $\delta$ which
originate from the induced four-gluon normalisation term in second order ([1])
contribute to the operator where
all external legs are attached to different vertices and therefore have to be
included
in the definitions:\\

$$\Pi^{\kappa\nu}_{AA}(x_1,x_2,\ldots):= t_{AA}^{\kappa\nu}(x_1,x_2,\ldots)+
\eqno(16)$$
$$-2\d_{\lambda}^{x_2}t^{\kappa\lambda\nu}_{AF}(x_1,x_2,\ldots)-2\d_{\lambda}^{x_1}t^{\lambda\kappa\nu}_{FA}(x_1,x_2,\ldots)+4\d_{\lambda}^{x_1}\d_{\tau}^{x_2}t^{\lambda\kappa\tau\nu}_{FF}(x_1,x_2,\ldots) $$
\\
$$\Pi^{\kappa l \nu}_{uA}(x_1,x_2,\ldots):= t^{\kappa l
\nu}_{uA}(x_1,x_2,\ldots)-2\d^{x_2}_{\lambda}t^{\kappa
l\lambda\nu}_{uF}(x_1,x_2,\ldots), l>2 $$
\\
$$\Pi^{\mu\nu}_{u\overline{u}A}(x_1,x_2,x_3,\ldots):=
t^{\mu\nu}_{u\overline{u}A}(x_1,x_2,x_3,\ldots) - 2 \d_{\kappa}^{x_3}
t^{\mu\kappa\nu}_{u\overline{u}F}(x_1,x_2,x_3,\ldots)$$
\\
$$\Pi^{\alpha\mu\nu}_{AAA}(x_1,x_2,x_3,\ldots):=
t^{\alpha\mu\nu}_{AAA}(x_1,x_2,x_3,\ldots)+$$ $$+2 g
\delta(x_1-x_2)t^{\nu\alpha\mu}_{AF}(x_3,x_2,\ldots)-2 g
\delta(x_1-x_3)t^{\mu\alpha\nu}_{AF}(x_2,x_3,\ldots)+2 g
\delta(x_2-x_3)t^{\alpha\mu\nu}_{AF}(x_1,x_2,\ldots)+$$
$$-2\d^{x_1}_{\kappa} [ t^{\kappa\alpha\mu\nu}_{FAA}(x_1,x_2,x_3,\ldots)+2 g
\delta(x_2-x_3)t^{\mu\nu\kappa\alpha}_{FF}(x_3,x_1,\ldots) ]+$$
$$-2 \d^{x_2}_{\kappa} [ t^{\alpha\kappa\mu\nu}_{AFA}(x_1,x_2,x_3,\ldots)-2 g
\delta(x_1-x_3)t^{\alpha\nu\kappa\mu}_{FF}(x_3,x_2,\ldots) ]+$$
$$-2\d^{x_3}_{\kappa} [ t^{\alpha\mu\kappa\nu}_{AAF}(x_1,x_2,x_3,\ldots)+2 g
\delta(x_1-x_2)t^{\kappa\nu\alpha\mu}_{FF}(x_3,x_2,\ldots) ]+$$
$$+4\d_{\kappa}^{x_2}\d_{\lambda}^{x_3}t^{\alpha\kappa\mu\lambda\nu}_{AFF}(x_1,x_2,x_3,\ldots)+4\d_{\kappa}^{x_1}\d_{\lambda}^{x_2}t^{\kappa\alpha\lambda\mu\nu}_{FFA}(x_1,x_2,x_3,\ldots)-8 \d_\kappa^{x_1} \d_\lambda^{x_2} \d_\sigma^{x_3}
t_{FFF}^{\kappa\alpha\lambda\mu\sigma\nu}(x_1,x_2,x_3,...)$$
\\
$$\Pi^{\alpha l\mu\nu}_{uAA}(x_1,x_2,x_3,\ldots):=t_{uAA}^{\alpha
l\mu\nu}(x_1,x_2,x_3,\ldots)+$$ $$-2\d^{x_3}_{\kappa}t^{\alpha
l\mu\kappa\nu}_{uAF}(x_1,x_2,x_3,\ldots)-2\d^{x_2}_{\kappa}t^{\alpha
l\kappa\mu\nu}_{uFA}(x_1,x_2,x_3,\ldots)+4\d^{x_2}_{\kappa}\d^{x_3}_{\lambda}t^{\alpha l\kappa\mu\lambda\nu}_{uFF}(x_1,x_2,x_3,\ldots)+$$ $$+2 g  \delta(x_2-x_3)t_{uF}^{\alpha (l-1)\mu\nu}(x_1,x_2,\ldots), l>3 $$
\\
$$\Pi^{\alpha\mu}_{Au \overline{u}}(x_1,x_2,x_3,.....) :=
t^{\alpha\mu}_{Au\overline{u}}(x_1,x_2,x_3,....)-2 \d_{\kappa}^{x_1}
t^{\kappa\alpha\mu}_{Fu\overline{u}} (x_1,x_2,x_3,...)$$
\\
$$\Pi^{\alpha\mu\nu}_{uA\overline{u}Aabcd}(x_1,x_2,x_3,x_4,...) :=
t^{\alpha\mu\nu}_{uA\overline{u}Aabcd}(x_1,x_2,x_3,x_4,....)+$$
$$ -2\d_{\kappa}^{x_4}
t_{uA\overline{u}Fabcd}^{\alpha \mu \kappa \nu}(x_1,x_2,x_3,x_4,....) -
2\d_{\kappa}^{x_2} t_{uF\overline{u}Aabcd}^{\kappa
\alpha\mu\nu}(x_1,x_2,x_3,x_4,...)+$$
$$+ 4 \d_{\kappa}^{x_2} \d_{\lambda}^{x_4}
t_{uF\overline{u}Fabcd}^{\kappa\alpha\mu\lambda\nu}(x_1,x_2,x_3,x_4,....) +
2 g f_{bdr} f_{acr} \delta(2-4)
t^{\mu\alpha\nu}_{u\overline{u}F}(x_1,x_3,x_4,.....)$$
\\
$$\bar\Pi^{3\nu}_{uu\overline{u}Aabcd}(x_1,x_2,x_3,x_4,...) := \bar
t^{3\nu}_{uu\overline{u}Aabcd}(x_1,x_2,x_3,x_4,...) -2 \d^{x_4}_{\alpha} \bar
t^{3\alpha\nu}_{uu\overline{u}Fabcd}(x_1,x_2,x_3,x_4,...)$$
\\
$$\Pi^{\alpha l \mu\nu}_{uu\overline{u}Aabcd}(x_1,x_2,x_3,x_4,...) := t^{\alpha
l \mu\nu}_{uu\overline{u}Aabcd}(x_1,x_2,x_3,x_4,...) - 2 \d^{x_4}_{\kappa}
t^{\alpha l \mu\kappa\nu}_{uu\overline{u}Fabcd}(x_1,x_2,x_3,x_4,...), l>4$$
\\
Analogously, one can define $\Pi^{\alpha\nu\kappa\lambda}_{AAAAabcd}$ and
$\Pi^{\alpha\kappa\lambda}_{u\overline{u}AAabcd}$.
\\
\\
$\bullet$ Having defined these summed distributions we arrive at the summed
two-leg identity by inserting (A.2) into (A.1) and using the new definitions.\\
$$ \d^{x_1}_{\kappa}\Pi^{\kappa\nu}_{AA}(x_1,x_2,x_3,\ldots
,x_{n-1})-\d^{\alpha}_{x_2}
[\d^{\alpha}_{x_2}t_{u\overline{u}}^{\nu}(x_1,x_2,x_3,\ldots
,x_{n-1})-(\alpha\leftrightarrow\nu)]+$$
$$+\sum^n_{l=3} \d_{\kappa}^{x_l}\Pi^{\kappa l\nu}_{uA}(x_1,x_2,x_3,\ldots
,x_{n-1})=0 \eqno(17)$$
Inserting (A.6) and (A.5) into (A.4) and using the new definitions, we arrive
at the first summed three-leg identities of gauge invariance:
$$ \d^{x_1}_{\alpha}\Pi^{\alpha\mu\nu}_{AAA}(x_1,x_2,x_3,x_4,\ldots ,x_n)+$$
$$+[\bigl(\d^{x_2}_{\alpha}[\d^{\alpha}_{x_2}\Pi^{\mu\nu}_{u\overline{u}A}(x_1,x_2,x_3,x_4,\ldots ,x_n)-(\alpha\leftrightarrow\mu)]\bigl) - \bigl((x_2,\nu)\longleftrightarrow(x_3,\mu)\bigl) ]+$$
$$+g[\delta(x_1-x_2)-\delta(x_1-x_3)]\Pi^{\mu\nu}_{AA}(x_2,x_3,x_4,\ldots
,x_n)+$$
$$+g[\bigl(\d^{x_2}_{\alpha}[\delta(x_2-x_3)g^{\alpha\mu}t^{\nu}_{u\overline{u}}(x_1,x_2,x_4,\ldots ,x_n)-(\alpha\leftrightarrow\nu)]\bigl) - \bigl((x_2,\nu)\longleftrightarrow(x_3,\mu)\bigl) ]+$$
$$+\sum^{n}_{l=4}\d^{l}_{\alpha}\Pi^{\alpha
l\mu\nu}_{uAA}(x_1,x_2,x_3,x_4,\ldots ,x_n) = 0 \eqno(18)$$
We can rewrite equation (A.3) as the second summed
three-leg identity:
$$\d_\alpha^{x_1}\Pi_{Au\tilde u}^{\alpha\mu}(x_1,x_2,x_3,\ldots)+
\d_\alpha^{x_2}\Pi_{Au\tilde u}^{\alpha\mu}(x_2,x_1,x_3,\ldots)+\d_{x_3}
^\mu\bar t_{uu\tilde u}^3(x_1,x_2,x_3,\ldots)+$$
$$+\sum_{l=4}^n\d_\alpha^{x_l}t_{uu\tilde u}^{\alpha l\mu}(x_1,x_2,x_3,
\ldots)+g\delta(x_1-x_2)t_{u\tilde u}^\mu(x_2,x_3,\ldots)+$$
$$-g\delta(x_1-x_3)t_{u\tilde u}^\mu(x_2,x_3,\ldots)-g\delta(x_2-x_3)
t_{u\tilde u}^\mu(x_1,x_3,\ldots)=0.\eqno(19)$$
Inserting (A.12) into (A.8), we get the first summed four-leg identity:
$$0= - \Bigl[\d_\alpha^{x_2}\Pi_{uA\tilde
uAabcd}^{\alpha\mu\nu}(x_1,x_2,x_3,x_4,\ldots) -\Bigl((a,x_1)\lra
(b,x_2)\Bigl)\Bigl]+$$
$$+\d_{x_3}^\mu\bar \Pi_{uu\tilde uAabcd}^{3\nu}(x_1,x_2,x_3,x_4,\ldots)
-\d_\alpha^{x_4}\Bigl[\d_\nu^{x_4} t_{uu\tilde u\tilde uabcd}^{\mu
\alpha}(x_1,x_2,x_3,x_4,\ldots)-(\nu\leftrightarrow\alpha)\Bigl]+$$
$$+\sum_{l=5}^n\d_\alpha^{x_l}\Pi_{uu\tilde uAabcd}^{\alpha
l\mu\nu}(x_1,x_2,x_3,x_4,\ldots)+$$
$$+g\{f_{abr}f_{cdr}[\delta(x_1-x_2)\Pi_{u\tilde
uA}^{\mu\nu}(x_2,x_3,x_4\ldots)+$$
$$-g\{f_{acr}f_{bdr}\delta(x_1-x_3)\Pi_{u\tilde
uA}^{\mu\nu}(x_2,x_3,x_4,\ldots)]-(a,x_1)\lra (b,x_2)\}+$$
$$+g\{f_{adr}f_{bcr}\delta(x_1-x_4)\Pi_{u\tilde
uA}^{\mu\nu}(x_2,x_3,x_4\ldots)-(a,x_1)\lra (b,x_2)\}+ $$
$$+g f_{abr}f_{cdr} \delta(x_3-x_4)g^{\nu\mu}\bar t_{uu\tilde
u}^3(x_1,x_2,x_4,\ldots)] \eqno(20)$$
Inserting (A.11) into (A.10), then (A.10) into (A.9) and finally (A.9) into
(A.7), we arrive at the second summed 4-leg identity :
$$
\d_\alpha^{x_1}\Pi_{AAAAabcd}^{\alpha\nu\kappa\lambda}(1,2,3,4,5,\ldots)+\Bigl\{ (-\d_\alpha^{x_2} \d^{\nu}_{x_2} \Pi_{u\overline{u}AAabcd}^{\alpha\kappa\lambda}(1,2,3,4,5, \ldots))-((\alpha\leftrightarrow\nu)\Bigl\}+$$
$$+\Bigl\{(b,\nu, x_2)\to (c,\kappa,x_3)\to (d,\lambda,x_4)\to (b,\nu,x_2)
\Bigl\}+\quad degenerate \quad terms =0 \eqno(21)$$
\\

These 5 (summed) identities are alike the Slavnov-Taylor identities. They can
directly compared with explicit identities which can be found in the
literature [5].  But
note that the summed cg-identities above are more general because in the usual
identities the adiabatic limit $g \rightarrow 1 \quad$in the inner variables is
already taken.\\
\\
Finally, we  derive the well-known relation between the Z-factors of the gluon
vertex, the gluon propagator, the ghost vertex and the ghost propagator at
one-loop level [5] from these summed Cg-identities :\\
$\bullet$ One easily checks that the following (local) renormalisations of the
self energy distributions  are compatible with the first summed identity of
gauge invariance (17) (and also with Lorentz invariance, all the discrete
symmetries and pseudo-unitarity) in the nth step of the inductive construction.
Because we are interested in the comparison with the Slavnov-Taylor identities,
we state only the relevant local normalisation terms which survive in the
adiabatic limit in regard to the inner coordinates:\\
$$ \Pi^{\mu\nu}_{AA} + C_{AA}^{n-1} [ \d^{\mu}_{x_1}\d^{\nu}_{x_1} -
g^{\mu\nu}\d_{x_1}\d_{x_1} ] \delta^{n-2} $$
$$ t^{\nu}_{u\overline{u}} + C_{u\overline{u}}^{n-1} \d^{\nu}_{x_2}
\delta^{n-2} \eqno(22)$$
$\bullet$The possible renormalisations of the two vertices (compatible with
Lorentz invariance,  discrete symmetries and pseudo-unitarity) are the
following:\\
$$ \Pi_{AAA}^{\alpha\mu\nu} + C_{AAA}^{n} [
g^{\alpha\mu}(\d^{\nu}_{x_1}-\d^{\nu}_{x_2})
+g^{\alpha\nu}(\d^{\mu}_{x_3}-\d^{\mu}_{x_1})
+g^{\mu\nu}(\d^{\alpha}_{x_2}-\d^{\alpha}_{x_3}) ] \delta^{n-1} $$
$$ \Pi_{u\overline{u}A}^{\mu\nu} + C_{u\overline{u}A}^{n}  \delta^{n-1}
g^{\mu\nu} \eqno(23)$$
The second summed identity (18) implies the following relation between these
four
normalisation constants in the nth step of the inductive construction:\\
$$ g C_{AA}^{n-1}+C_{u\overline{u}A}^{n}- g C_{u\overline{u}}^{n-1}-C_{AAA}^{n}
= 0 \eqno(24) $$
Because of $Z_{i}:=1+C_{i}$ (Note our conventions in the ghost sector!), we
directly get the well-known relation between the Z-factors at one-loop level:\\
$$\frac{Z_{AA}}{Z_{AAA}} = \frac{Z_{u\overline{u}}}{Z_{u\overline{u}A}}
\eqno(25) $$

The interpretation of this relation is slightly different in the causal
approach: It represents the restrictions by gauge invariance on finite
normalisation terms only. We do not need any infinite part in the Z-factors to
absorb divergences in the causal approach.\\
Using the two summed 4-leg identities or the identities, one can analogously
deduce the corresponding relation of the Z-factor of the four-gluon vertex.\\
\\

\vskip1cm

{\large\bf Appendix  $\quad$ Cg-Identities}
\vskip1cm

The conventions of denoting operator-valued distributions are as in [1]:
$$t_{AB\ldots ab\ldots}^{\alpha 2}(x_1,x_2,\ldots):A^a(x_1)B^b(x_2)\ldots:$$
means an operator-valued  distribution with external field operators (legs)
$A^a$ and
$B^b$, $a$ and $b$ are colour indices. The subscripts $\alpha 2$ show
that this term belongs to $T^\alpha_{n/2}(x_1,x_2,\ldots)$ with
$Q$-vertex at the second argument of the numerical distribution $t$.
All 2-leg distributions contain the colour tensor $\delta_{ab}$, all 3-leg
distributions the colour factor $f_{abc}$. Therefore we define the numerical
distributions without these colour factors . \\
\\

For $:\Omega:=\delta_{ab}:u_a(x_1)A_\nu^b(x_2):$ we obtain
$$\d_\alpha^1 t_{AA}^{\alpha\nu}+{1\over 2}\d_\alpha^2
[t_{uA}^{\alpha 2\nu}-t_{uA}^{\nu 2\alpha}]
+\sum_{l=3}^n\d_\alpha^l t_{uA}^{\alpha l\nu}=0,\eqno(A.1)$$

For $:\Omega:=\delta_{ab}:u_a(x_1) F_{\mu \nu}^b (x_2):$
$$\d_\alpha^1 t_{AF}^{\alpha\mu\nu}+{1\over 2}[\d_2^\mu
t_{u\tilde u}^{\nu}-\d_2^\nu t_{u\tilde u}^\mu]
+{1\over 4}[t_{uA}^{\mu 2\nu}-t_{uA}^{\nu 2\mu}]+
\sum_{l=3}^n\d_\alpha^l t_{uF}^{\alpha l\mu\nu}=0.\eqno(A.2)$$

For $:\Omega:=f_{abc}:u^a(x_1)u^b(x_2)\d_\mu\tilde u
^c(x_3):$
$$0=\d_\alpha^{x_1}t_{Au\tilde u}^{\alpha\mu}(x_1,x_2,x_3,\ldots)+
\d_\alpha^{x_2}t_{Au\tilde u}^{\alpha\mu}(x_2,x_1,x_3,\ldots)+\d_{x_3}
^\mu\bar t_{uu\tilde u}^3(x_1,x_2,x_3,\ldots)$$
$$+\sum_{l=4}^n\d_\alpha^{x_l}t_{uu\tilde u}^{\alpha l\mu}(x_1,x_2,x_3,
\ldots)+g\delta(x_1-x_2)t_{u\tilde u}^\mu(x_2,x_3,\ldots)$$
$$-g\delta(x_1-x_3)t_{u\tilde u}^\mu(x_2,x_3,\ldots)-g\delta(x_2-x_3)
t_{u\tilde u}^\mu(x_1,x_3,\ldots).\eqno(A.3)$$
For $:\Omega:=f_{abc}:u^a(x_1)A_\mu^b(x_2)A_\nu^c(x_3):$
$$0=\d_\alpha^{x_1}t_{AAA}^{\alpha\mu\nu}(x_1,x_2,x_3,x_4,\ldots)-
{1\over 2}\d_\alpha^{x_2}\Bigl[t_{uAA}^{\alpha 3\nu\mu}(x_1,x_3,x_2,x_4,
\ldots)-(\alpha\leftrightarrow\mu)\Bigl]$$
$$+{1\over 2}\d_\alpha^{x_3}\Bigl[t_{uAA}^{\alpha 3\mu\nu}(x_1,x_2,x_3,
x_4,\ldots)-(\alpha\leftrightarrow\nu)\Bigl]+\sum_{l=4}^n\d_\alpha^{x_l}
t_{uAA}^{\alpha l\mu\nu}(x_1,x_2,x_3,x_4,\ldots)$$
$$+g[\delta(x_1-x_2)-\delta(x_1-x_3)]t_{AA}^{\mu\nu}(x_2,x_3,x_4,\ldots)$$
$$-g\delta(x_2-x_3){1\over 2}\Bigl[t_{uA}^{\mu
2\nu}(x_1,x_2,x_4,\ldots)-(\mu\leftrightarrow\nu)\Bigl].\eqno(A.4)$$
For $:\Omega:=f_{abc}:u^a(x_1)A_\mu^b(x_2)F_{\nu\lambda}
^c(x_3):$
$$0={1\over 4}\Bigl[t_{uAA}^{\nu 3\mu\lambda}(x_1,x_2,x_3,\ldots)-(\nu
\leftrightarrow\lambda)\Bigl]$$
$$+\d_\alpha^{x_1}t_{AAF}^{\alpha\mu\nu\lambda}(x_1,x_2,x_3,\ldots)+{1\over
2}\d_\alpha^{x_2}\Bigl[t_{uAF}^{\alpha
2\mu\nu\lambda}(x_1,x_2,x_3,\ldots)-(\alpha\leftrightarrow\mu)\Bigl]$$
$$+{1\over 2}\Bigl[\d_{x_3}^\nu t_{u\tilde uA}^{\lambda\mu}(x_1,x_3,x_2,
\ldots)-(\nu\leftrightarrow\lambda)\Bigl]+\sum_{l=4}^n\d_\alpha^{x_l}t_{uAF}^{\alpha l\mu\nu\lambda}(x_1, x_2,x_3,\ldots)$$
$$+g[\delta(x_1-x_2)-\delta(x_1-x_3)]t_{AF}^{\mu\nu\lambda}(x_2,x_3,x_4,\ldots)$$
$$+{g\over 2}\Bigl[g^{\mu\nu}\delta(x_2-x_3)t_{u\tilde u}^\lambda(x_1,
x_3,x_4,\ldots)-(\nu\leftrightarrow\lambda)\Bigl].\eqno(A.5)$$
For $:\Omega:=f_{abc}:u^a(x_1)F_{\mu\tau}^b(x_2)F_{\nu\lambda}^c(x_3):$
$$0={1\over 4}\Bigl[t_{uAF}^{\mu
2\tau\nu\lambda}(x_1,x_2,x_3\ldots)-(\mu\leftrightarrow\tau)\Bigl]-{1\over
4}\Bigl[t_{uAF}^{\nu 2\lambda
\mu\tau}(x_1,x_3,x_2\ldots)-(\nu\leftrightarrow\lambda)\Bigl]$$
$$+\d_\alpha^{x_1}t_{AFF}^{\alpha\mu\tau\nu\lambda}(x_1,x_2,x_3,\ldots)+{1\over
2}\Bigl[\d_{x_2}^\tau t_{u\tilde u
F}^{\mu\nu\lambda}(x_1,x_2,x_3\ldots)-(\mu\leftrightarrow\tau)\Bigl]$$
$$-{1\over 2}\Bigl[\d_{x_3}^\lambda t_{u\tilde uF}^{\nu\mu\tau}(x_1,
x_3,x_2\ldots)-(\lambda\leftrightarrow\nu)\Bigl]$$
$$+\sum_{l=4}^n\d_\alpha^{x_l}t_{uFF}^{\alpha
l\mu\tau\nu\lambda}(x_1,x_2,x_3\ldots)$$
$$+g[\delta(x_1-x_2)-\delta(x_1-x_3)]t_{FF}^{\mu\tau\nu\lambda}(x_2,
x_3,x_4\ldots).\eqno(A.6)$$

$:\Omega:=:u^a(x_1) A_\nu^b(x_2) A_\kappa^c(x_3)$ $A_\lambda^d(x_4):$
$$0 =
\d_\alpha^{x_1}t_{AAAAabcd}^{\alpha\nu\kappa\lambda}(1,2,3,4,5,\ldots)+\eh\d_\alpha^{x_2}[t_{uAAAabcd}^{\alpha 2\nu\kappa\lambda}(1,2,3,4,5, \ldots)-(\alpha\leftrightarrow\nu)]$$
$$+\eh\d_\alpha^{x_3}[t_{uAAAacbd}^{\alpha 2\kappa\nu\lambda}(1,3,2,4,5,
\ldots)-(\alpha\leftrightarrow\kappa)]+\eh\d_\alpha^{x_4}[t_{uAAAadbc}^ {\alpha
2\lambda\nu\kappa}(1,4,2,3,5,\ldots)-(\alpha\leftrightarrow\lambda)]$$
$$+\sum_{l=5}^n\d_\alpha^{x_l}t_{uAAAabcd}^{\alpha
l\nu\kappa\lambda}(1,2,3,4,5, \ldots)+$$
$$+\Bigl\{gf_{abr}f_{cdr}\Bigl[\delta(1-2)t_{AAA}^{\nu\kappa\lambda}
(2,3,4,5,\ldots)+\delta(3-4)\eh
[t_{uAA}^{\kappa 2\lambda\nu}(1,3,2,5,\ldots)-(\kappa\leftrightarrow
\lambda)]\Bigl]\Bigl\}$$
$$+\Bigl\{(b,\nu, x_2)\to (c,\kappa,x_3)\to (d,\lambda,x_4)\to (b,\nu,x_2)
\Bigl\},\eqno(A.7)$$

For $:\Omega:=:u^a(x_1)u^b(x_2)\d_\mu\tilde u^c(x_3)A_\nu^d(x_4):$
$$0= - \Bigl[\d_\alpha^{x_2}t_{uA\tilde
uAabcd}^{\alpha\mu\nu}(1,2,3,4,5,\ldots) -\Bigl((a,x_1)\lra
(b,x_2)\Bigl)\Bigl]$$
$$+\d_{x_3}^\mu\bar t_{uu\tilde uAabcd}^{3\nu}(1,2,3,4,5,\ldots)
+\d_\alpha^{x_4}\eh\Bigl[t_{uu\tilde uAabcd}^{\alpha 4\mu\nu}(1,2,3,4,5,
\ldots)-(\alpha\leftrightarrow\nu)\Bigl]$$
$$+\sum_{l=5}^n\d_\alpha^{x_l}t_{uu\tilde uAabcd}^{\alpha l\mu\nu}(1,2,3,4,5,
\ldots)$$
$$+g\{f_{adr}f_{bcr}[\delta(1-4)t_{Au\tilde u}^{\nu\mu}(4,2,3,5\ldots)$$
$$+\delta(2-3)t_{Au\tilde u}^{\nu\mu}(4,1,3,5\ldots)]\}-g\{(a,x_1)\lra
(b,x_2)\}$$
$$+gf_{abr}f_{cdr}[\delta(1-2)t_{Au\tilde u}^{\nu\mu}(4,2,3,5\ldots)$$
$$+\delta(3-4)g^{\nu\mu}\bar t_{uu\tilde u}^3(1,2,4,5\ldots)],
\eqno(A.8)$$

For $:\Omega:=:u^a(x_1)F_{\kappa\lambda}^b(x_2)A_\mu ^c(x_3)A_\nu^d(x_4):$
$$0=\d_\alpha^{x_1}t_{AFAAabcd}^{\alpha\kappa\lambda\mu\nu}(1,2,3,4,5,\ldots)
+\eh[\d_{x_2}^\lambda t_{u\tilde uAAabcd}^{\kappa\mu\nu}(1,2,3,4,5,\ldots)
-(\lambda\lra\kappa)]$$
$$+\eh\d_\alpha^{x_3}[t_{uFAAabcd}^{\alpha 3\kappa\lambda\mu\nu}
(1,2,3,4,5,\ldots)-(\alpha\lra\mu)]$$
$$+\eh\d_\alpha^{x_4}[t_{uFAAabdc}^{\alpha 3\kappa\lambda\nu\mu}(1,2,4,3,5,
\ldots)-(\alpha\leftrightarrow\nu)]
+\sum_{l=5}^n\d_\alpha^{x_l}t_{uFAAabcd}^{\alpha l\kappa\lambda\mu\nu}
(1,2,3,4,5,\ldots)$$
$$+{\scriptstyle{1\over 4}}[t_{uAAAabcd}^{\kappa 2\lambda\mu\nu}(1,2,3,4,5,
\ldots)-(\kappa\leftrightarrow\lambda)]$$
$$+gf_{abr}f_{cdr}\delta(3-4)\eh [t_{uAF}^{\mu 2\nu\kappa\lambda}
(1,3,2,5\ldots)-(\mu\lra\nu)]$$
$$+{g\over 2}\{f_{adr}f_{bcr}\delta(2-3)[g^{\mu\kappa}t_{u\tilde uA}
^{\lambda\nu}(1,3,4,5\ldots)-(\kappa\lra\lambda)]\}+{g\over 2}\{(c,\mu, 3)\lra
(d,\nu, 4)\}$$
$$+g[f_{acr}f_{dbr}\delta(1-3)-f_{adr}f_{cbr}\delta(1-4)]t_{AAF}^
{\mu\nu\kappa\lambda}(3,4,2,5\ldots)$$
$$+gf_{abr}f_{cdr}\delta(1-2)t_{AAF}^{\mu\nu\kappa\lambda}(3,4,2,5\ldots)],\eqno(A.9)$$

For $:\Omega:=:u^a(x_1)A_\mu^b(x_2)F_{\kappa\lambda}^c(x_3)F_{\sigma
\tau}^d(x_4):$
$$0=\d_\alpha^{x_1}t_{AAFFabcd}^{\alpha\mu\kappa\lambda\sigma\tau}
(1,2,3,4,5,\ldots)+\eh\d^{x_2}_\alpha[t_{uAFFabcd}^{\alpha 2\mu\kappa
\lambda\sigma\tau}(1,2,3,4,5,\ldots)-(\alpha\lra\mu)]$$
$$+\eh\{\d^\lambda_{x_3}t_{uA\tilde uFabcd}^{\mu\kappa
\tau\rho}(1,2,3,4,5,\ldots)-(\kappa\lra\lambda)]\}$$
$$+\eh\{(c,\kappa,\lambda,x_3)\lra (d,\sigma,\tau,x_4)\}
+\sum_{l=5}^n\d_\alpha^{x_l}t_{uAFFabcd}^{\alpha l\mu\kappa\lambda\sigma
\tau}(1,2,3,4,5,\ldots)$$
$$+{\scriptstyle{1\over 4}}[t_{uAAFabcd}^{\kappa 3\mu\lambda\sigma\tau}
(1,2,3,4,5,\ldots)-(\kappa\leftrightarrow\lambda)]+{\scriptstyle{1\over 4}}
[(c,\kappa,\lambda,x_3)\lra (d,\sigma,\tau,x_4)]$$
$$+g\{f_{abr}f_{cdr}\delta(1-2)t_{AAF}^{\mu\kappa\lambda\sigma\tau}
(2,3,4,5\ldots)$$

$$-{g\over 2}\{f_{adr}f_{bcr}\delta(2-3)[g^{\mu\kappa}t_{u\tilde uF}
^{\lambda\sigma\tau}(1,3,4,5\ldots)-(\kappa\lra\lambda)]\} $$
$$-{g\over 2}\{(c,\kappa,\lambda,x_3)\lra (d,\sigma,\tau,x_4)\}$$
$$+g[f_{adr}f_{bcr}\delta(1-4)-f_{acr}f_{bdr}\delta(1-3)]t_{AFF}^
{\mu\kappa\lambda\sigma\tau}(2,3,4,5\ldots),\eqno(A.10)$$

For $:\Omega:=:u^a(x_1)F_{\mu\nu}^b(x_2)F_{\kappa\lambda}
^c(x_3)F_{\sigma\tau}^d(x_4):$
$$0=\d_\alpha^{x_1}t_{AFFFabcd}^{\alpha\mu\nu\kappa\lambda\sigma\tau}
(1,2,3,4,5,\ldots)+\eh[\d_{x_2}^\nu t_{u\tilde uFFabcd}^{\mu\kappa
\lambda\sigma\tau}(1,2,3,4,5,\ldots)-(\mu\lra\nu)]$$
$$+\eh[\d_{x_3}^\lambda t_{u\tilde u FFacbd}^{\kappa\mu\nu\sigma\tau}
(1,3,2,4,5,\ldots)-(\kappa\lra\lambda)]$$
$$+\eh[\d_{x_4}^\tau t_{u\tilde u FFadbc}^{\sigma\mu\nu\kappa\lambda}
(1,4,2,3,5,\ldots)-(\sigma\leftrightarrow\tau)]
+\sum_{l=5}^n\d_\alpha^{x_l}t_{uFFFabcd}^{\alpha l\mu\nu\kappa\lambda
\sigma\tau}(1,2,3,4,5,\ldots)$$
$$+{\scriptstyle{1\over 4}}[t_{uAFFabcd}^{\mu 2\nu\kappa\lambda\sigma\tau}
(1,2,3,4,5,\ldots)-(\mu\leftrightarrow\nu)] +{\scriptstyle{1\over
4}}[t_{uAFFacbd}^{\kappa 2\lambda\mu\nu \sigma\tau}
(1,3,2,4,5,\ldots)-(\kappa\leftrightarrow\lambda)]$$
$$+{\scriptstyle{1\over 4}}[t_{uAFFadbc}^{\sigma 2\tau\mu\nu\kappa\lambda}
(1,4,2,3,5, \ldots)-(\sigma\leftrightarrow\tau)]$$
$$-g[f_{acr}f_{bdr}\delta(1-3)+f_{adr}f_{cbr}\delta(1-4)
+f_{abr}f_{dcr}\delta(1-2)]t_{FFF}^{\mu\nu\kappa\lambda\sigma\tau}
(2,3,4,5\ldots),\eqno(A.11)$$

For $:\Omega:=:u^a(x_1)u^b(x_2)\d_\mu\tilde u^c(x_3)F_{\lambda\kappa}^ d(x_4):$
$$0=\Bigl[\d_\alpha^{x_1}t_{Au\tilde uFabcd}^{\alpha\mu\lambda\kappa}
(1,2,3,4,5,\ldots)-\Bigl((a,x_1)\lra (b,x_2)\Bigl)\Bigl]$$
$$+\d_{x_3}^\mu\bar t_{uu\tilde uFabcd}^{3\lambda\kappa}(1,2,3,4,5,\ldots)
+\eh\Bigl[\d_{x_4}^\kappa t_{uu\tilde u\tilde u abcd}^{\mu\lambda}
(1,2,3,4,5,\ldots)-(\kappa\leftrightarrow\lambda)\Bigl]$$
$$+\sum_{l=5}^n\d_\alpha^{x_l}t_{uu\tilde uFabcd}^{\alpha l\mu\lambda
\kappa}(1,2,3,4,5,\ldots)$$
$$+{\scriptstyle{1\over 4}}[t_{uu\tilde uAabcd}^{\lambda 4\mu\kappa}
(1,2,3,4,5,\ldots)-(\lambda\leftrightarrow\kappa)]$$
$$+gf_{abr}f_{cdr}\delta(1-2)t_{u\tilde uF}^{\mu\lambda\kappa}
(2,3,4,5\ldots)$$
$$-g[f_{acr}f_{bdr}\delta(1-3)t_{u\tilde uF}^{\mu\lambda\kappa}(2,3,4,5
\ldots)-((a,x_1)\lra (b,x_2))]$$
$$+g[f_{adr}f_{bcr}\delta(1-4)t_{u\tilde uF}^{\mu\lambda\kappa}(2,3,4,5
\ldots)-((a,x_1)\lra (b,x_2))],\eqno(A.12)$$

\vskip2cm

{\large\bf Acknowledgements}
\vskip0.3cm
I thank A.Aste and M.D\"utsch for useful discussions and the Swiss National
Science Foundation for financial support.

\vskip2cm

{\large\bf References}

\vspace{0.5cm}

\begin{tabbing}

1. \quad\quad\quad\= M. D\"utsch, T. Hurth, G. Scharf,\\
 \>   Nuovo Cimento 108A (1995) 679, 108A (1995) 737 \\
 \> T. Hurth,\\
 \>  Annals of Physics (1995) to appear, hep-th/9411080\\
2. \> H. Epstein, V. Glaser, \\
 \> Annales de l'Institut Poincare 29 (1973) 211\\
 \> G. Scharf, \\
 \> Finite Quantum Electrodynamics (Second Edition),\\
 \>  Texts and Monographs in Physics , Springer (1995) to appear\\
3. \> T. Hurth,\\
 \>  Z\"urich University Preprint ZU-TH-20/95\\
4. \> M. D\"utsch,\\
 \>  Z\"urich University Preprint ZU-TH-10/95\\
5. \> P. Pascual, R. Tarrach \\
 \> Nuclear Physics B174 (1980) 123\\
 \> S.K. Kim, M. Baker, \\
 \> Nuclear Physics B164 (1980) 152\\
 \>  T.W. Chiu,\\
 \>  Nuclear Physics B181 (1980) 450\\
 \> (and references therein)
\end{tabbing}

\vspace{0.5cm}

\newpage

\newpage

\end{document}